# Barrier Inhomogeneity of Schottky Diode on Nonpolar AlN Grown by Physical Vapor Transport

Qin Zhou, Honglei Wu, Hui Li, Xi Tang, *Member IEEE*, Zuoyan Qin, Dan Dong, Yan Lin, Chengjin Lu, Ran Qiu, Ruisheng Zheng, Jiannong Wang, and Baikui Li, *Member IEEE*

*Abstract*—An aluminum nitride (AlN) Schottky barrier diode (SBD) was fabricated on a nonpolar AlN crystal grown on tungsten substrate by physical vapor transport. The Ni/Au–AlN SBD features a low ideality factor $n$ of 3.3 and an effective Schottky barrier height (SBH) of 1.05 eV at room temperature. The ideality factor $n$ decreases and the effective SBH increases at high temperatures. The temperature dependences of $n$ and SBH were explained using an inhomogeneous model. A mean SBH of 2.105 eV was obtained for the Ni–AlN Schottky junction from the inhomogeneity analysis of the current–voltage characteristics. An equation in which the parameters have explicit physical meanings in thermionic emission theory is proposed to describe the current-voltage characteristics of inhomogeneous SBDs.

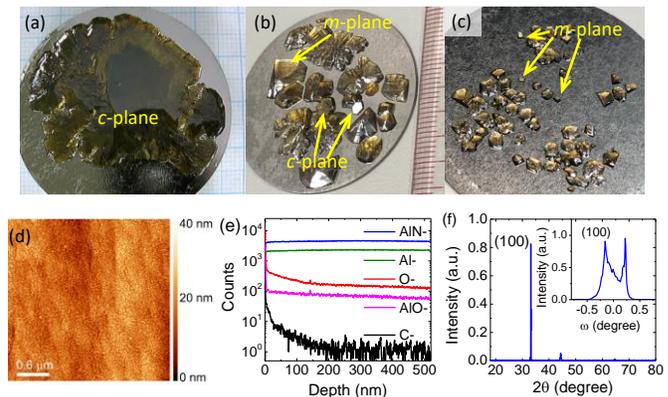

Fig. 1. (a)-(c) *c*-plane and *m*-plane AlN crystals grown by physical vapor transport on tungsten substrates through spontaneous nucleation. (d) Surface morphology, (e) SIMS profile, (f) XRD pattern, of the AlN crystal used in this work. The inset of (f) shows the rocking curve of the (100) plane.

## I. INTRODUCTION

Aluminum nitride (AlN), with a bandgap energy of 6.1 eV and a critical electric field of up to 12 MV/cm, possesses great potential in power electronics and deep UV optoelectronics [1]–[3]. Though reported almost a century earlier [4] than its counterpart, gallium nitride (GaN) [5], which has been the main force of power electronics [6], [7] and optoelectronics [8], [9], high-quality AlN crystals and epilayers had not been achieved until recently [10]–[15]. AlN's large energy bandgap results in a high activation energy for *n*/*p* dopants [2] and a wide accommodation energy range for mid-gap defects or traps [4], [16]. Schottky barrier diodes (SBDs) on AlN grown by physical vapor transport (PVT) [17], by hydride vapor phase epitaxy (HVPE) [18], and by metal organic chemical vapor deposition (MOCVD) [19] have been reported. An AlN metal–semiconductor field-effect transistor (MESFET) on MOCVD-grown epilayer was also demonstrated recently [20]. These SBDs and MESFET, though with performances far from the predicted material's limit, indicate the feasibility of developing AlN electronic devices.

The ideality factor $n$ and the Schottky barrier height (SBH) are the two critical parameters used to evaluate the quality of an SBD [21], [22]. Compared with GaN [23], [24] and SiC [25], [26] SBDs, of which the ideality factors near a unity have been achieved, the reported $n$ of AlN SBDs is still much larger [17]–[19]. Due to the ultrawide bandgap and the large value of the charge neutral level (CNL) of AlN [27], [28], a larger SBH has always been expected for AlN SBDs. By performing the X-ray photoelectron spectroscopy (XPS) measurements, Reddy *et al*. reported SBHs of 1.6 eV to 2.3 eV (2.2 eV to 2.4 eV) for different metals on *m*-plane (*c*-plane) AlN [29]. However, the SBHs extracted from the current–voltage (*J–V*) curves of fabricated AlN SBDs were much smaller, ~1.0 eV [17]–[19]. The underlying mechanism of the large discrepancy in the SBHs of AlN SBDs extracted from the different measurements is still unclear.

In this work, we reported on a Ni/AlN SBD fabricated on a nonpolar AlN crystal grown on tungsten (W) substrate by PVT. The temperature dependences of the ideality factor $n$ and the

This work was supported by National Natural Science Foundation of China under Grant 61604098 and by Shenzhen Science and Technology Innovation Commission under Grant 20170302143001451 and Grant 20160520174438578 and also in part by the Research Grants Council of the Hong Kong SAR under Grant C6013-16E. (*Q. Zhou and H. Wu contributed equally to this work*) (*Corresponding author: Baikui Li*).

Q. Zhou, H. Wu, X. Tang, Z. Qin, D. Dong, Y. Lin, C. Lu, R. Qiu, R. Zheng, and B. Li are with the College of Physics and Optoelectronic Engineering, Shenzhen University, Shenzhen 518060, China. (e-mail: libk@szu.edu.cn).

H. Li and J. N. Wang are with the Department of Physics, the Hong Kong University of Science and Technology, Clear Water Bay, Hong Kong, China.

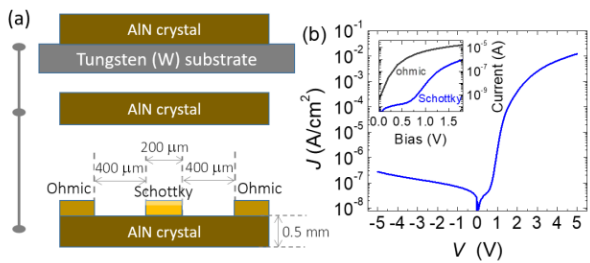

Fig. 2. (a) Schematic cross-sections showing the process flow of the Ni/AlN SBD. (b) $J$–$V$ curve at room temperature. The inset shows the current–voltage characteristics of the ohmic contacts and that of the Schottky contact.

effective SBH extracted from the $J$–$V$ curves were explained using an inhomogeneous model [22], [30]–[33]. A mean SBH of 2.105 eV was obtained for the Ni–AlN Schottky junction. A capacitance–voltage ($C$–$V$) measurement was also performed.

## II. GROWTH, FABRICATION, AND CHARACTERIZATIONS

The AlN was grown by PVT on a tungsten (W) substrate through spontaneous nucleation [15]. Figure 1(a)–1(c) show the optical views of the typical $c$-plane and $m$-plane AlN crystals grown on W substrates by PVT under different nucleation conditions. With an optimized temperature gradient on the W substrate, AlN crystals over a centimeter in size can be achieved. In this work, an AlN crystal with a rectangular shape of 1.6 mm × 2.5 mm and a thickness of approximately 0.5 mm was used. Figure 1(d) illustrates the surface morphology of the AlN crystal examined by atomic force microscopy (AFM). The root-mean-square (RMS) roughness of a 3 μm × 3 μm area is approximately 3 nm. Figure 1(e) shows the secondary ion mass spectrometry (SIMS) result of the AlN crystal measured using negative polarity mode. Oxygen is the dominant bulk impurity. The existence of carbon at the surface is due to surface contamination. The nonpolar surface was identified to be the (100) plane via the X-ray diffraction (XRD) pattern of the AlN crystal, as shown in Fig. 1(f). The splitting of the (100) peak in the rocking curve, as shown in the inset of Fig. 1(f), reveals that the AlN crystal employed here is not an ideal single crystal.

Figure 2(a) summarizes the fabrication processes and the schematic cross-section of the Ni/AlN SBD. The W substrate was first etched using HF:HNO$_3$ solution. The AlN crystal was cleaned in acetone and isopropyl alcohol under ultrasonic radiation for removal of surface contaminations and in diluted alkaline solutions to remove the native oxidation [34]. For ohmic contacts, a Ti/Al/Ni/Au (20 nm /150 nm /50 nm /200 nm) metal stack was first deposited by e-beam evaporation and then annealed at 900 ºC for 30 s in N$_2$ atmosphere using rapid thermal annealing (RTA). For the Schottky contact, a Ni/Au (20 nm /200 nm) metal stack was deposited and then annealed at 350 °C for 5 min in N$_2$ atmosphere to improve the interface quality of the contact [9], [35], [36].

Figure 2(b) plots the $J$–$V$ curve of the Ni/AlN SBD at 300 K. The $J$–$V$ curve can be described by the empirical equation [21],

$$J = J_0 \left[\exp\left(\frac{qV}{nkT}\right) - 1\right], \text{ and } J_0 = A^*T^2 \exp\left(-\frac{q\phi_{B,(0,T)}^{J-V}}{kT}\right) \quad (1)$$

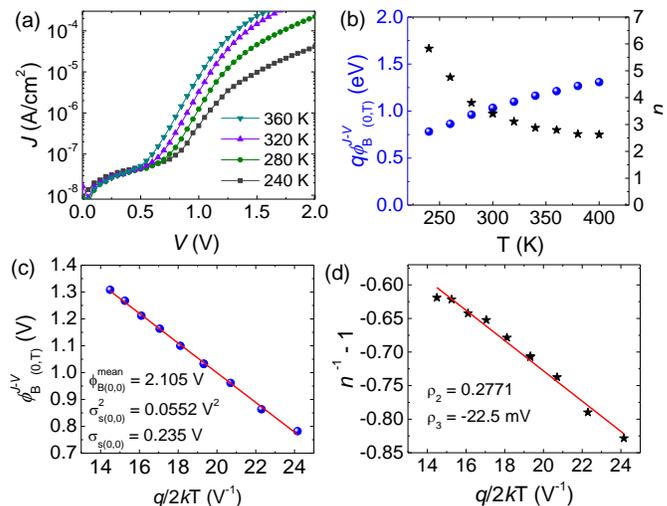

Fig. 3. (a) $J$–$V$ curves, (b) Extracted $n$ and the effective SBH, at different temperatures of the Ni/AlN SBD. Plots of (c) $\phi_{B,(0,T)}^{J-V}$, and (d) ($n^{-1}$–1) as functions of $q/2kT$. The solid lines are fitting curves using the inhomogeneity model, with the fitting results illustrated in the figure.

where $A^*$ is the Richardson constant, $T$ is temperature, $n$ is the ideality factor, and $q\phi_{B,(0,T)}^{J-V}$ is the SBH at zero bias. The subscript (0,T) is used because the SBH here is calculated from $J_0$, which is the intercept at zero bias of the extrapolation of the linear segment of the (ln$J$)–$V$ curve at temperature $T$. Variations in the $A^*$ value in a relatively wide range around the theoretical value have little influence on the extraction of SBH using (1) [37]. Taking the value of $A^*$ to be ~57.6 A/(cm$^2$K$^2$) for AlN [18], the ideality factor $n$ and SBH of the Ni/AlN SBD at 300 K were extracted to be 3.3 and 1.05 eV, respectively, from the linear segment of the ln($J$)-$V$ curve. As shown in the inset of Fig. 2(b), although the ohmic contacts on the AlN are not ideal, the resistance between the two ohmic contacts is several orders smaller than that of the Schottky contact, and its effect on the extraction of junction parameters from the linear segment of the (ln$J$)-$V$ curve is negligible. A higher annealing temperature and/or contact metal with lower work function are required to realize ideal ohmic contacts on the unintentionally doped AlN crystal. The reverse leakage current of this bare device is relatively high. Optimized device structures including proper passivation and field-plate design should be introduced to enhance the reverse bias blocking capability

## III. JUNCTION INHOMOGENEITY ANALYSIS

Figure 3(a) shows the $J$–$V$ curves of the Ni/AlN SBD at different temperatures. The extracted $n$ and effective SBH as functions of the temperature are plotted in Fig. 3(b). As the temperature increases from 240 K to 400 K, the $n$ decreases from 5.8 to 2.6, and the effective SBH increases from 0.8 eV to 1.3 eV. These temperature dependences indicate that the Ni–AlN Schottky interface is inhomogeneous [22], [30]–[33].

Assuming that the local Schottky barriers at the Ni–AlN interface follow a Gaussian distribution $P(\phi_B)$ with a standard deviation of $\sigma$ around a mean value of $\phi_B^{mean}$ [22], [30], we have,

$$P(\emptyset_B) = \frac{1}{\sigma_{(V,0)}\sqrt{2\pi}} \exp\left[-\frac{(\emptyset_{B,(V,0)}^{mean} - \emptyset_B)^2}{2\sigma_{(V,0)}^2}\right],$$

$$\text{and} \int_{-\infty}^{+\infty} P(\emptyset_B)\, d\emptyset_B = 1 \quad (2)$$

Here, the barrier potential fluctuation or distribution is assumed to be independent of temperature but may be affected by the applied biases due to image-force lowering [21]. The current density from metal to semiconductor in thermionic emission theory can be written as,

$$J_{ms} = A^*T^2 \exp\left(-q\emptyset_{B,(V,T)}^{J-V}/kT\right)$$
$$= A^*T^2 \int_{-\infty}^{+\infty} \exp(-q\emptyset_B/kT)\, P(\emptyset_B) d\emptyset_B \quad (3)$$

By integrating (3), the equation

$$\emptyset_{B,(V,T)}^{J-V} = \emptyset_{B,(V,0)}^{mean} - \frac{\sigma_{(V,0)}^2}{2kT/q} \quad (4)$$

can be obtained. For an ideal SBD, the SBH decreases slightly at high temperatures due to the temperature dependence of the semiconductor's energy bandgap [21], [22]. For an inhomogeneous SBD, however, the effective SBH described by (4) increases at high temperatures. The underlying physical mechanism of this temperature dependence can be revealed by dividing the inhomogeneous interface into a high SBH region and a low SBH region. The ratio of the junction current through the high SBH region to that through the low SBH region in thermionic emission theory is $\exp(-q(\emptyset_{B,high} - \emptyset_{B,low})/kT)$, which increases at high temperatures, consequently leading to an increase in the effective SBH.

With a temperature- and bias-dependent SBH, the $J-V$ of an SBD in thermionic emission theory can be written as,

$$J = A^*T^2 \exp\left(-\frac{q\emptyset_{B,(V,T)}^{J-V}}{kT}\right)\left[\exp\left(\frac{qV}{kT}\right) - 1\right] \quad (5)$$

Equating (1) and (5), the ideality factor $n$ in the empirical equation can be obtained with the following,

$$\frac{1}{n} - 1 = -\frac{\emptyset_{B,(V,T)}^{J-V} - \emptyset_{B,(0,T)}^{J-V}}{V} \quad (6a)$$

For a typical SBD of which the $J-V$ curve follows (1), $n$ is constant as the inverse of the slope of the linear segment of the $\ln(J)-V$ curve. Consequently, the dependence of $\emptyset_{B,(V,T)}^{J-V}$ on $V$ should be linear. At temperature $T$, (6a) can be written as,

$$n^{-1}(T) - 1 = -\frac{\emptyset_{B,(V,T)}^{J-V} - \emptyset_{B,(0,T)}^{J-V}}{V} = -\frac{\Delta\emptyset_{B,(V,T)}^{J-V}}{V} = -\frac{\rho_1(T)V}{V}$$
$$= -\rho_1(T) \quad (6b)$$

Here, $\rho_1$ is a temperature-dependent coefficient. According to (4), the bias-induced change in SBH can be written as,

$$\Delta\emptyset_{B,(V,T)}^{J-V} = \Delta\emptyset_{B,(V,0)}^{mean} - \frac{\Delta\sigma_{(V,0)}^2}{2kT/q}$$

with $\Delta\emptyset_{B,(V,0)}^{mean} = \rho_2 V$, and $\Delta\sigma_{(V,0)}^2 = \rho_3 V \quad (7)$

where $\rho_2$ and $\rho_3$ are temperature-independent coefficients, representing the bias dependences of the mean value and the standard deviation of the inhomogeneous junction, respectively. Combining (6b) and (7), it can be obtained that,

$$n^{-1}(T) - 1 = -\rho_1(T) = -\rho_2 + \frac{\rho_3}{2kT/q} \quad (8)$$

then, the physical meaning of the ideality factor $n$ in the widely used empirical equation is revealed as that it represents the voltage deformation of the barrier distribution of an inhomogeneous Schottky junction.

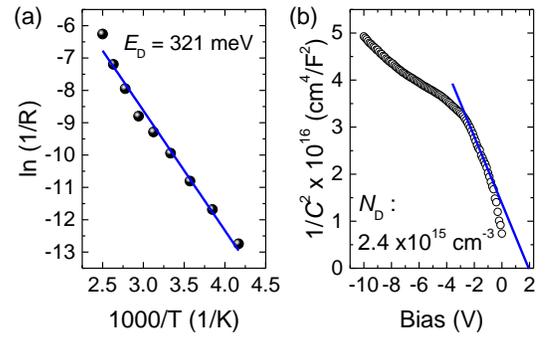

Fig. 4. (a) Arrhenius plot of $1/R$, where $R$ is the differential resistance at 10 V between the two ohmic contacts. (b) $1/C^2-V$ plot of the $C-V$ curve measured at 1 kHz at room temperature of the Ni/AlN SBD.

Figure 3(c) and 3(d) plot the extracted $q\emptyset_{B,(0,T)}^{J-V}$ and $n^{-1}-1$ of the Ni/AlN SBD as functions of $q/2kT$, respectively. A linear fitting of the data in Fig. 3(c) using (4) gives the mean value of the SBH, $q\emptyset_{B,(0,0)}^{mean}$, as 2.105 eV with a standard deviation, $q\sigma_{(0,0)}$, of 0.235 eV at zero bias. A linear fitting of the data in Fig. 3(d) using (8) gives the voltage dependences of the mean SBH and the standard deviation with the coefficients $\rho_2 = 0.277$ and $\rho_3 = -22.5$ mV, respectively.

The mean SBH is in agreement with the CNL and the predicted SBH values of AlN [27]–[29]. The positive value of $\rho_2$ indicates that the mean SBH increases with forward biases. The negative value of $\rho_3$ indicates that the standard deviation of the inhomogeneous Schottky junction decreases with forward biases, i.e., the junction becomes effectively more uniform under forward biases. These results are consequential when considering the image-force lowering at the metal–semiconductor interface [21], [22]. Besides, in an inhomogeneous Schottky junction, the low SBH region is usually "pinched-off" by the surrounding high SBH region, leading to a pronounced band bending in the low SBH region [38], which dominates the transport behaviors of the inhomogeneous SBD. Under forward biases, numerical simulation has shown that the maximum point of the bent conduction band rises and also shifts from the interface towards the bulk region of the semiconductor [21], [22], [38], resulting consequently in an increase of the effective SBH and also a relatively more uniform Schottky interface. The mixed surface polarity or the existence of crystal domain boundaries (Fig. 1(f)) and the surface roughness (Fig. 1(d)) may lead to the inhomogeneity of the SBH.

The as-grown AlN is $n$-type [17], [18]. Although the resistance of the ohmic contact on the unintentionally doped AlN is much smaller than that of the Schottky contact, reliable Hall measurements could not be conducted to measure the electron mobility and density. To estimate the electron density and the background dopant density in the AlN crystal grown on W substrate by PVT, we assumed that the electron mobility was constant in the temperature range studied in this work. Then, the inverse of the resistance $(1/R)$ can be described as $1/R \sim$ $\exp(-E_D/kT)$, where $E_D$ is the donor activation energy [2], [18]. As shown in Fig. 4(a), an $E_D$ of 321 meV can be extracted. It should be pointed out that this activation energy is underestimated by approximately 10% because the electron mobility increases at low temperatures [2], [18]. Substitutional oxygen $O_N$ is the possible dopant in the unintentionally-doped PVT-grown AlN [17], [39]. The effective donor concentration $N_D$ was estimated to be $2.4 \times 10^{15}$ cm$^{-3}$ from the slope of $1/C^2$–$V$ [19] at low reverse biases, as shown in Fig. 4(b). At high reverse biases, the value of the capacitance was affected by the leakage current, which is relatively high for the bare SBD. The electron density $n_e$ was estimated to be $2 \times 10^{14}$ cm$^{-3}$ from the simplified charge-neutrality equation, $n_e^2 = (N_D N_C/2)\exp(-E_D/kT)$, where $N_C$ is the effective density of states in the conduction band [18]. Then the energy difference $E_F$ between the Fermi level and the conduction band minimum was found to be 273 meV from the equation $n_e = N_C \exp(-E_F/kT)$. The unintentionally-doped AlN crystals grown by PVT are usually insulating [18], [40]. One possible reason for the relatively high conductivity of the AlN crystal grown by PVT on W substrate in this work is the absence of carbon, which introduces compensation doping to nitride semiconductors [41], [42].

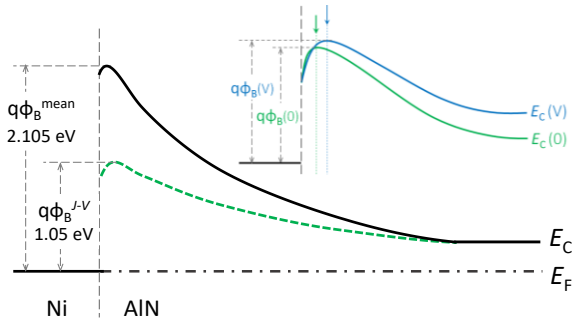

Fig. 5. Schematic band diagram of the inhomogeneous Ni/AlN Schottky barrier diode, with the mean SBH of the barrier distribution and the effective SBH of $J$-$V$ characteristics illustrated. The inset illustrates schematically the bias deformation of the effective Schottky barrier height, i.e., the maximum point of the bent conduction band becomes higher and moves away from the interface towards the AlN bulk.

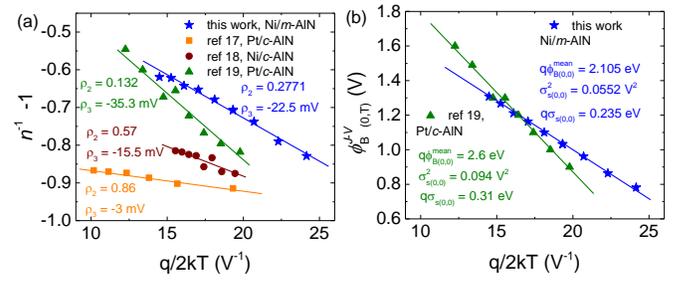

Fig. 6. Replots of, (a) the ideality factors and, (b) the effective SBH, reported in references as a function of $q/2kT$. The solid lines are fitting curves using the inhomogeneity model. The fitting results are also shown in the figure.

The SBH can also be obtained from the $C$-$V$ measurement [21]. Because the capacitance is determined mainly by the bulk properties of the depletion region of the semiconductor, the interface inhomogeneity plays little role in the extraction of SBH from the $C$-$V$ curve [30]. When the total diode capacitance under reverse bias was modeled as a parallel combination of elemental capacitances, taking the inhomogeneity of the depletion width determined by the local barrier height into account, the value of SBH extracted from the $C$–$V$ measurement is slightly larger than that extracted from the $J$–$V$ analysis [43]. As shown in Fig. 4(b), the intercept $V_0$ on the voltage axis of the $1/C^2$–$V$ curve is about 1.95 V. Using the equation $q\Phi_B^{C-V} = qV_0 + kT + E_F$, the $q\Phi_B^{C-V}$ was obtained to be 2.249 eV, which is slightly larger than the mean SBH (i.e., 2.105 eV) obtained from the inhomogeneity analysis of the $J$–$V$ characteristics. This result verifies the validity of analyzing the $J$–$V$ characteristics of the AlN SBDs using the inhomogeneity model.

The inhomogeneity analysis explains the significant deviation of the low effective SBH in $J$–$V$ characteristics of state-of-the-art AlN SBDs from the predicted SBH which is much larger due to the ultrawide bandgap and large CNL value of AlN. Figure 5 illustrates schematically the band diagrams of an Ni/AlN SBD. The inset of Fig. 5 illustrates schematically the bias deformation of the effective SBH. Under forward biases, the maximum point of the bent conduction band becomes higher and moves away from the interface towards the AlN bulk, i.e., the effective SBH increases and Schottky interface becomes effectively less inhomogeneous. The $J$–$V$ behaviors of the inhomogeneous Ni/AlN SBD are dominated by the low SBH region. With advanced material qualities and surface/interface engineering techniques to suppress the formation of the low SBH region, AlN SBDs with larger effective SBHs in $J$–$V$ characteristics are procurable for high voltage rectification applications.

## IV. Discussions

Figures 6(a) and 6(b) replot the ideality factors and effective SBHs of the different AlN SBDs that have been reported so far as functions of $q/2kT$, respectively. All the data can be well fitted using the inhomogeneity model, indicating that the inhomogeneity analysis is generally applicable to the ultrawide bandgap AlN SBDs. For the Pt/$c$-AlN SBD reported in Ref. 19, the fitted value of the mean SBH is 2.6 eV, which is larger than that of the Ni/$m$-AlN SBD, i.e., 2.105 eV, reported in this work. Besides the difference of metal work functions, this result indicates a lower CNL on the $c$-plane than that on the $m$-plane of AlN, in agreement with the XPS studies by Reddy *et. al.* [27], [29].

The physical meaning of the ideality factor $n$ in the widely used empirical equation is more obvious in the form of $(n^{-1}-1)$, which represents the voltage deformation of an inhomogeneous Schottky junction. Here, we introduce a factor $m = (n^{-1}-1)$. Then the equation describing the $J$–$V$ of an inhomogeneous SBD can be re-written as follows, in which the parameters have explicit physical meanings in thermionic emission theory,

$$J = A^*T^2 \exp\left(-\frac{q(\varnothing_{B,(0,T)}^{J-V} + mV)}{kT}\right)\left[\exp\left(\frac{qV}{kT}\right) - 1\right],$$
$$\text{where } 0 < m < 1 \qquad (9)$$

The barrier distribution and voltage deformation of the inhomogeneous Schottky junction can be determined from the temperature dependences of $\varnothing_{B,(0,T)}^{J-V}$ and $m$, respectively.